\documentclass[sigconf]{acmart}

\setcopyright{none}
\renewcommand\footnotetextcopyrightpermission[1]{}

\usepackage{fancyhdr}
\fancypagestyle{plain}{%
   \fancyhf{} %
   \fancyfoot[L]{Technical report}%
   \fancyfoot[R]{Submitted on December 2, 2021}%
}
\fancypagestyle{firstpagestyle}{%
   \fancyhf{} %
   \fancyfoot[L]{Technical report}%
   \fancyfoot[R]{
   	 Submitted on December 2, 2021
   }%
}

\usepackage{balance}

\usepackage{booktabs}

\usepackage{xspace}

\usepackage{todonotes}

\newcommand{\picoquic}{\texttt{picoquic}\xspace}
\newcommand{\pquic}{\texttt{PQUIC}\xspace}
\newcommand{\spns}{single packet number space\xspace}
\newcommand{\mpns}{multiple packet number spaces\xspace}

\usepackage{caption}
\usepackage{subcaption}
\usepackage[htt]{hyphenat}

\begin{document}

\title{The Packet Number Space Debate in Multipath QUIC}

\author{Quentin De Coninck}
\authornote{FNRS Post-doctoral Researcher.}
\affiliation{
	\institution{UCLouvain, Belgium}
}
\email{quentin.deconinck@uclouvain.be}

%
%
%

\begin{abstract}
	With a standardization process that attracted many interest, QUIC can been seen as the next general-purpose transport protocol.
	Still, it does not provide true multipath support yet, missing some use cases that MPTCP can address.
	To fill that gap, the IETF recently adopted a multipath proposal merging all the proposed designs. 
	While it focuses on its core components, there still remains one major design issue in the proposal: the number of packet number spaces that should be used.
	This paper provides experimental results with two different Multipath QUIC implementations based on NS3 simulations to understand the impact of using one packet number space per path or a single packet number space for the whole connection.
	Our results suggest that using one packet number space per path makes the Multipath QUIC connection more resilient to the receiver's acknowledgment strategy.
\end{abstract}

%



\maketitle

\section{Introduction}

QUIC is a recently standardized protocol~\cite{rfc9000} aiming at providing the services of TLS/TCP (built-in encryption, reliable data transfer,...) with data multiplexing atop UDP.
Initially designed for HTTP/3~\cite{ietf-quic-http-34}, QUIC sets itself up as the next general purpose transport protocol for the Internet and many extensions, such as the support for unreliable data transfer~\cite{ietf-quic-datagram-06}, have been proposed.
Thanks to its flexibility, QUIC can serve many use cases~\cite{kosek2021beyond} while enabling rapid experiments with, e.g., congestion control algorithms~\cite{kakhki2017taking,narayan2018restructuring}.


While QUIC supports probing new networks and switching to a different network, it only provides single-path data transmission.
In particular, endhosts cannot simultaneously use different network paths to, e.g., aggregate their  bandwidths.
Still, there is demand for such real multipath support for various use cases~\cite{interim-20-10}.
Multipath TCP~\cite{raiciu2012hard} and CMT-SCTP~\cite{iyengar2006concurrent} can now address them, such as mobility support for delay-sensitive applications~\cite{de2018tuning}, 
network handovers in high-speed trains~\cite{li2018measurement} and hybrid access networks~\cite{bonaventure2016multipath}.


However, both Multipath TCP and CMT-SCTP faced deployment issues on the Internet~\cite{honda2011still,budzisz2012taxonomy}.
QUIC mitigates such network interference by design thanks to its built-in encryption, raising interest to bring multipath support.
The first attempts~\cite{de2017multipath, viernickel2018multipath} were mostly built on the design experience of Multipath TCP.
However, these were based on an old version of QUIC~\cite{langley2017quic} which differs from the standardized one.
As of September 2021, there were three different multipath proposals for standardized QUIC: \texttt{draft-deconinck-quic-multipath}~\cite{deconinck-quic-multipath-07,de2021multiflow}, \texttt{draft-huitema-quic-mpath-option}~\cite{huitema-quic-mpath-option-01} and \texttt{draft-liu-multipath-quic}~\cite{liu-multipath-quic-04,zheng2021xlink}.
Having several proposals actually slowed down reaching the consensus on one approach.
Some were considered too complex.
For instance, \texttt{draft-deconinck} provides support for unidirectional paths (while QUIC's path validation ensures bidirectional ones) and IP address communication (raising concerns about forged addresses).
\texttt{draft-liu} adds Quality of Experience signaling, which might not be used in some use cases.

In order to advance the multipath work, all the previous drafts' authors worked together to make a common proposal~\cite{lmbdhk-quic-multipath-00} focusing on the core components that would suit all the aforementioned use cases.
This multipath draft recently got adopted at IETF112~\cite{ietf112-minutes}.
However, there remains a core design issue that requires consensus: the amount of packet number spaces that a multipath QUIC connection should use.
While \texttt{draft-huitema} advocates for keeping a single application packet number space, \texttt{draft-deconinck} and \texttt{draft-liu} dedicate an application packet number space per path.
While the unified proposal enables negotiating both options, it is likely only one will be widely supported in the end.

This paper aims at providing insights to the network research community in order to understand the implications of this design choice, not only for Multipath QUIC, but also for any multipath transport protocol.
Indeed, MPTCP has several path's TCP sequence numbers with a global MPTCP data sequence one, while CMT-SCTP has a sequence number per data stream.
Our evaluation reveals that while both designs work for QUIC, using a \spns makes the performance of the transfer more dependent on the receiver's acknowledgment strategy than using multiple ones.

The remaining of this paper is organized as follows.
We start in Section~\ref{sec:multipath} by describing the core components of Multipath QUIC and explaining the advantages and drawbacks of each packet number space design.
Then, in Section~\ref{sec:evaluation}, we evaluate these designs by considering two different implementations (\picoquic~\cite{huitema_picoquic_2021} and \pquic~\cite{de2019pluginizing}) under a broad range of network scenarios using the NS3 simulator~\cite{riley2010ns}.
Finally, we discuss our results in Section~\ref{sec:discussion}. 

\section{Bringing Multipath to QUIC}\label{sec:multipath}

QUIC packets are fully encrypted, except a small header containing a few clear-text fields.
Among them, the \emph{Destination Connection ID} enables the endhosts to map packets to QUIC connections.
This makes QUIC unbound to the 4-tuple (IP\textsubscript{src}, IP\textsubscript{dst}, port\textsubscript{src}, port\textsubscript{dst}).
Some servers negotiate 0-length Connection IDs to limit the wire overhead.
In such cases, the QUIC connection is identified by the 4-tuple.
QUIC packets contain a monotonically increasing packet number that is used to compute a unique AEAD nonce of at least 64 bits.
When a packet is lost, its content can be retransmitted but in a packet with a greater packet number.
While SCTP has chunks, the core QUIC data, located in the encrypted payload, are called \emph{frames}.
The \texttt{STREAM} frame carries the application data.
This frame contains an absolute offset that does not wrap-around, unlike TCP's sequence number and MPTCP's Data Sequence Number.
Furthermore, unlike (MP)TCP and (CMT-)SCTP, QUIC does not acknowledge data sequence numbers, but packets numbers using the \texttt{ACK} frame.
Compared to TCP where the number of selective acknowledgments~\cite{mathis1996rfc2018} is often limited to 2-3 per packet, an \texttt{ACK} frame is theoretically constrained to the size of the QUIC packet, limiting the number of ranges to several hundreds entries.
In addition, the ACK delay field of the \texttt{ACK} frame includes the time between the reception of the largest packet number and the instant the frame is sent, allowing precise RTT estimations.

One of the key features of QUIC is connection migration.
For instance, a connection initiated by a smartphone can move from a Wi-Fi network to a cellular one due to, e.g., user mobility.
For that, a client willing to change the network path needs first to check that the endpoint is still reachable using the new 4-tuple before initiating connection migration.
This process is called \emph{path validation}.
To do so, both endhosts need to have an unused Destination Connection ID provided by their peer through \texttt{NEW\_CONNECTION\_ID} frames\footnote{Unless the peer negotiated 0-length Connection IDs.}.
An endhost starts the path validation process by sending a packet containing a \texttt{PATH\_CHALLENGE} frame containing 8 bytes of opaque data from a new 4-tuple with an unused Destination Connection ID.
Upon reception of such packet, the peer replies on the newly-seen 4-tuple with another 
packet containing a \texttt{PATH\_RESPONSE} frame echoing the opaque data of the \texttt{PATH\_CHALLENGE} frame using an unused Destination Connection ID.
Upon reception of that packet, the endhost marks the path as validated.
At this point, the peer may also initiate path validation on this new path.
As soon as the client starts sending data (e.g. \texttt{STREAM} or \texttt{ACK} frames) over this new network path, the server stops using the previous path and migrates the connection over the new 4-tuple.

While QUIC bundles such connection migration feature, it does not enables hosts to simultaneously use several network paths to send data.
The unified proposal~\cite{lmbdhk-quic-multipath-00} strives at providing multipath usage with as few changes as possible to the QUIC specification~\cite{rfc9000}.
It focuses on the core building blocks: negotiating multipath, initiating new paths and numbering packets.
Advanced multipath-specific algorithms, such as packet scheduling and selecting the paths to use, are out-of-scope of the proposal.

\begin{figure*}
	\centering
	\begin{subfigure}[b]{0.495\textwidth}
		\centering
		\includegraphics[width=.77\columnwidth]{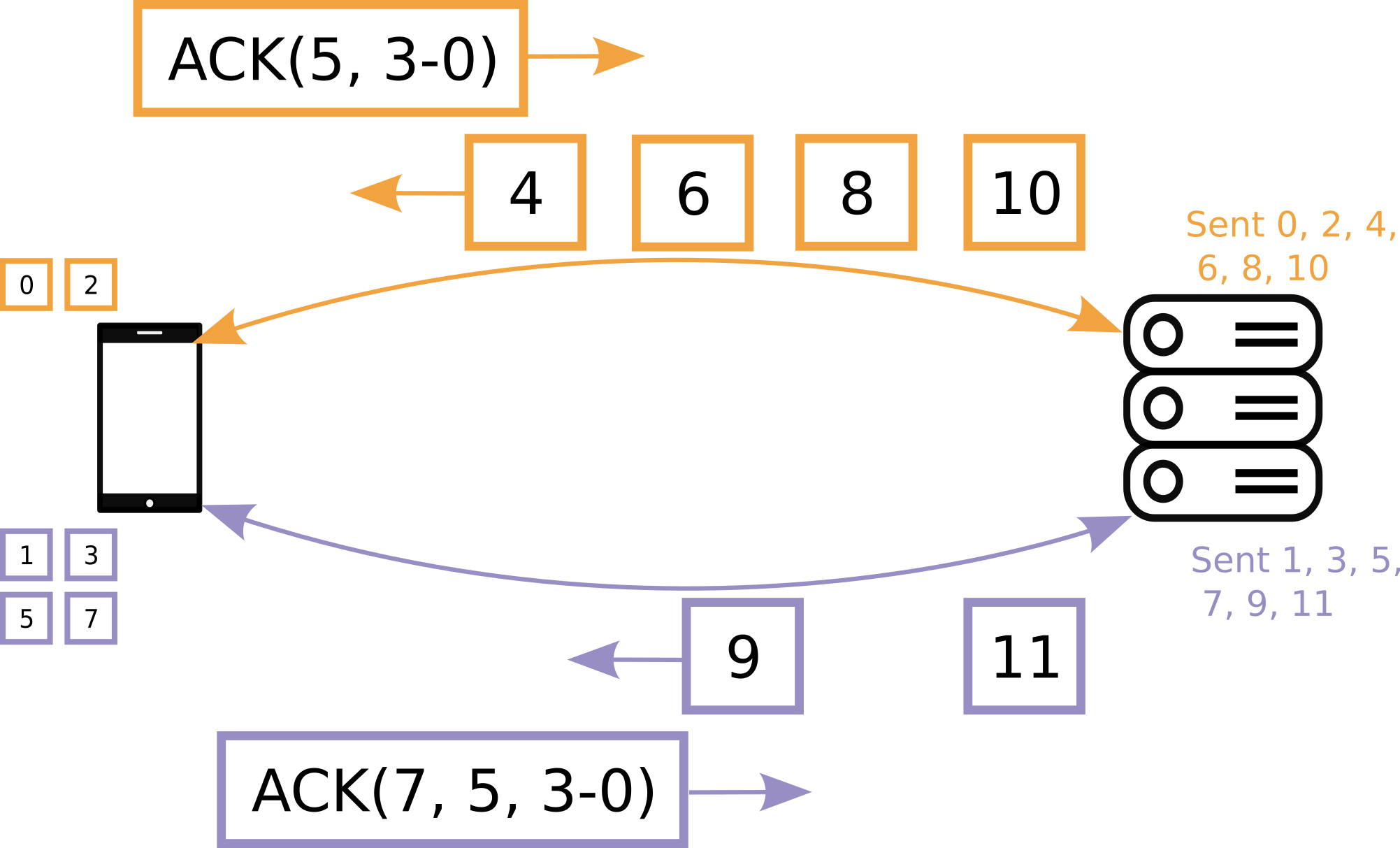}
		\caption{Single packet number space case. }
		\label{fig:spns}
	\end{subfigure}
	\hfill
	\begin{subfigure}[b]{0.495\textwidth}
		\centering
		\includegraphics[width=.77\columnwidth]{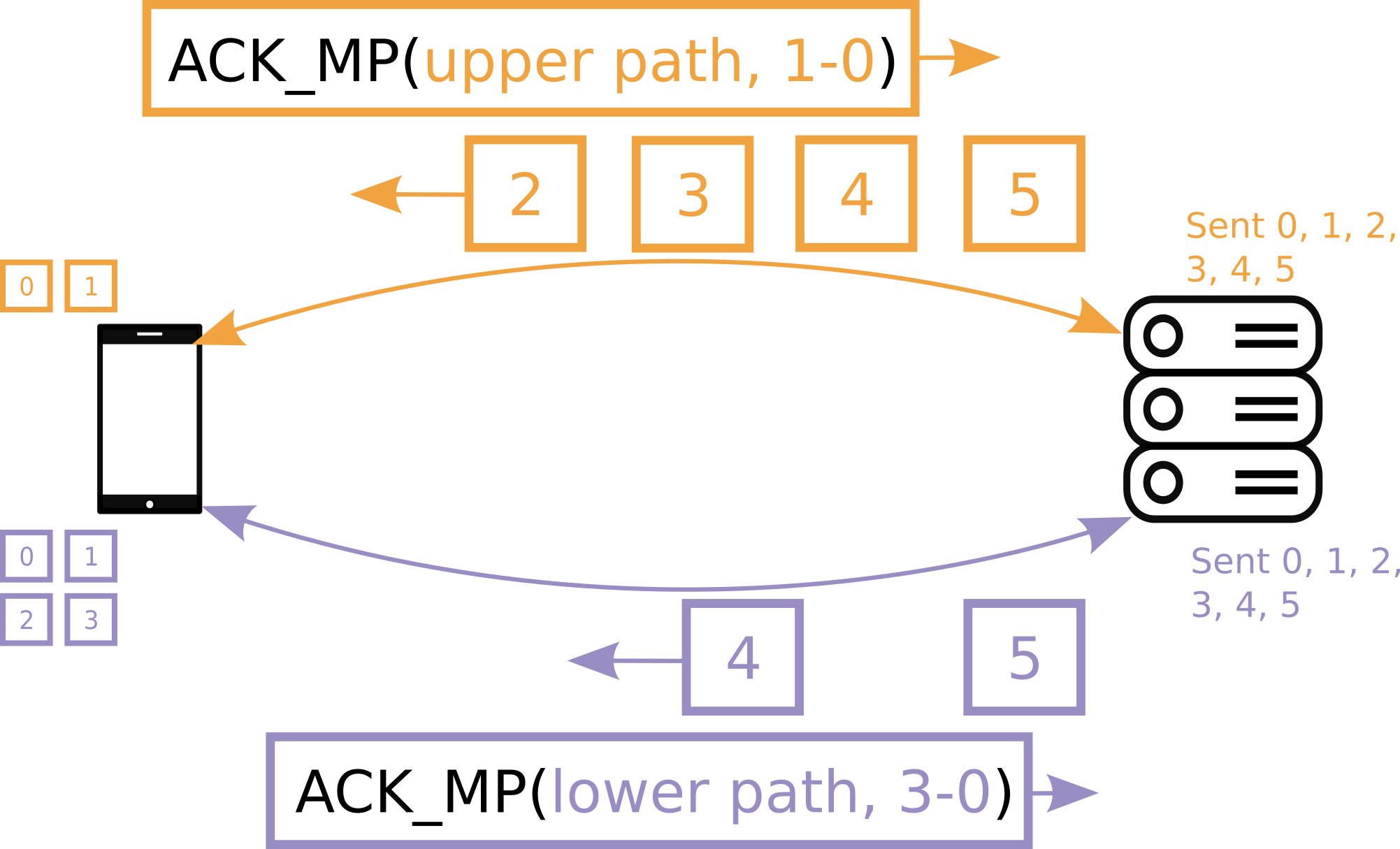}
		\caption{Multiple packet number spaces case.}
		\label{fig:mpns}
	\end{subfigure}
	\hfill
	\caption{An example of a 12-packets data transfer where the server follows a round-robin scheduling strategy to send packets to the client. It shows a snapshot where the fourth packet on the bottom path got received. For the sake of this example, the client replies with a packet containing an acknowledgment frame on the same path than the last received packet, but in practice the client has no constraint on the path it should use.}
	\label{fig:example}
\end{figure*}

\paragraph{Multipath Negotiation.}
As many other QUIC extensions, the multipath one is negotiated during the connection handshake using a QUIC transport parameter, named \texttt{enable\_multipath}, to negotiate its usage.
If both the client and the server advertise compatible values indicating multipath support\footnote{The multipath negotiation allows hosts to advertise to advertise their support for one of the packet number space design, or both. If, e.g., the client has exclusive support for \spns while the server only handles \mpns, then the multipath negotiation will fail.}, then the connection uses the multipath extension.
Note that the handshake needs to be fully completed before starting using multiple paths.

\paragraph{Initiating New Paths.}
Multipath QUIC builds on the path validation process to initiate new paths.
The main addition to the single-path QUIC is that with the multipath extension, several validated paths can be simultaneously used to send data packets.
Each path, using different 4-tuples, is identified by the sequence number of the Destination Connection ID (communicated by the NEW CONNECTION ID frame) it uses.
Packets can then be mapped to a path thanks to the packet's Connection ID, or the packet's 4-tuple when zero-length Connection IDs are used. 
Note that endhosts can change at any time the Connection ID used over a given path.
Furthermore, since an endhost must use distinct Destination Connection IDs per path, its peer controls the maximum number of paths that can be opened by choosing when to send \texttt{NEW\_CONNECTION\_ID} frames\footnote{A hard limit on the number of usable Connection IDs is determined during the handshake thanks to the \texttt{active\_connection\_id\_limit} QUIC transport parameter.}.


\paragraph{Numbering Packets.}
In single-path QUIC, all post-handshake packets use a single application packet number space.
The current multipath proposal enables implementations to experiment with either single packet number space or multiple packet number spaces.
We first describe the \spns design.
Then, we discuss the \mpns one.

\subsection{Single Packet Number Space}

This design lets endhosts spread packets over different paths while keeping the regular QUIC application packet number space.
This means that a packet with number $N$ can be sent over path $A$ while a packet with number $N+1$ can be transmitted on path $B$.
To illustrate this situation, Figure~\ref{fig:spns} represents a server sending 12 packets to a client in a round-robin fashion.
Here, the server sends all even packet numbers on the upper path and all the odd ones on the lower path.
The packets are acknowledged with the regular ACK frame.
However, when paths do not exhibit the same performance, the receiver is likely to observe out-of-order  packet number reception.
In the example, the lower path is faster than the upper one.
When the client receives packet number 2 on the upper path, it advertised that it received the two first packets on the upper path (0, 2) and the three first packets on the lower one (1, 3, 5) by sending an ACK frame with two ranges: 5 and from 3 to 0.
Then, it receives the fourth packet on lower path (7).
This creates a new range inside the ACK frame.
Note that the receiver cannot know \emph{a priori} the next packet number it should expect on a path.

The main concern about this approach relates to the number of ranges that the \texttt{ACK} frame sent by the receiver could contain.
When paths have very different latencies, the receiver should pay attention not to prune its ranges too quickly.
Furthermore, as we will see in Section~\ref{sec:evaluation}, implementations might want to limit the number of ranges they encode in \texttt{ACK} frames to limit the state they need to maintain.
In such cases, some packets might be either lately or never acknowledged by the receiver, even if they were finally received.
In addition, the RTT estimates are also less precise because the ACK delay field of the \texttt{ACK} frame only relates to the largest packet number received.
This makes the performance of a multipath transfer more dependent on the receiver's strategy.

\subsection{Multiple Packet Number Spaces}

With this design, each path has an associated packet number space.
As depicted in Figure~\ref{fig:mpns}, subsequent packets over a given path use consecutive packet number spaces.
Because packet numbers are now path-dependent, an augmented version of the ACK frame, called the \texttt{ACK\_MP} frame, includes the path identifier to which the acknowledged packet numbers refer to.
Such an approach avoids large ranges in \texttt{ACK\_MP} frames due to different paths' performances, making the multipath performance less dependent of the receiver's strategy.
It also enables keeping a simple per-path logic for lost packet detection.

Yet, the \mpns design also has drawbacks.
First, it requires endhosts to use non-zero-length Connection IDs to be able to identify the packet number space of a packet.
Second, it involves changes to the use of AEAD for packet protection.
QUIC uses the packet number to compute the AEAD nonce, and this nonce must not be reused over a given connection.
To mitigate this issue when using \mpns, the nonce includes the path identifier. 

\section{Evaluation}\label{sec:evaluation}

To evaluate the impact of the number of packet number spaces on Multipath QUIC performance, we explore a large set of network scenarios within the NS3 environment~\cite{riley2010ns}.
Compared to emulation, this setup enables fully reproducible and stable results while still using actual implementations.

We focus on two different open-source implementations of Multipath QUIC.
\picoquic~\cite{huitema_picoquic_2021} supports both single packet number space and multiple packet number spaces following the unified proposal.
We want to emphasize that all the algorithms (loss detection, multipath-specific decisions,...) remain the same regardless of the packet number space design used.
\pquic~\cite{de2019pluginizing} has a multipath plugin implementing an earlier proposal~\cite{deconinck-quic-multipath-07} relying on multiple packet number spaces.
Note that both implementations limit to 32 the maximum number of additional ACK Blocks (AB) contained in a single \texttt{ACK(\_MP)} frame, therefore limiting the maximum number of ranges within an \texttt{ACK(\_MP)} frame to 33.
We start our experiments with \picoquic\footnote{Commit \texttt{f4ae862}.} and \pquic\footnote{Commit \texttt{841c822}.} versions that acknowledge first the ranges that are the closest to latest acknowledged packet number.

We explore many network situations following an experimental design approach using the WSP algorithm~\cite{santiago_construction_2012} enabling us to cover broadly the parameter space with 95 points, resulting in 95 runs.
All our experiments consist in a 50 MB download initiated by a GET request over a single stream.
Relying on such a large transfer, combined to a large initial receive buffer of 2 MB enables us to limit the impact of the packet scheduler.
The client initiates the usage of all the available paths upon handshake completion.
We define the transfer time as the delay between the first QUIC packet sent by the client and the last QUIC packet received by the client containing the \texttt{CONNECTION\_CLOSE} frame.
In addition, we generate QLOG files~\cite{marx2020debugging} at client and server sides to get a full view of the connection and the internal state of the implementations.

We first experiment with 2-path network scenarios where both paths share the same characteristics.
Then, we explore 2-path situations where paths exhibit heterogeneous performances in terms of bandwidth and delay.
Finally, we extend our heterogeneous scenarios to 3-path networks.

\subsection{Homogeneous 2-Path Experiments}

\begin{table}
	\centering
	\caption{Parameter space for the 95 homogeneous runs.}
	\begin{tabular}{@{}c|cc@{}}
		\toprule
		Factor & Min & Max \\
		\midrule
		Bandwidth [Mbps] & 2.5 & 100 \\
		RTT [ms] & 5 & 100 \\
		\bottomrule
	\end{tabular}
	\label{tab:homo_param_space}
\end{table}

When all the network paths provide the same performance, i.e., same delay and bandwidth, there should not be many packet reordering at receiver's side.
Therefore, there should not be much performance difference between packet number space designs. 
To assess this intuition, we consider the parameter space depicted in Table~\ref{tab:homo_param_space}.
In this paper, unless explicitly mentioned otherwise, we only consider loss-less networks and the buffer size of the bottleneck router is always set to 1.5 times the bandwidth-delay product.
Note that actual packet losses may still occur due to bottleneck router's buffer overflow.


\begin{figure*}
	\centering
	\begin{subfigure}[b]{0.33\textwidth}
		\centering
		\includegraphics[width=\columnwidth]{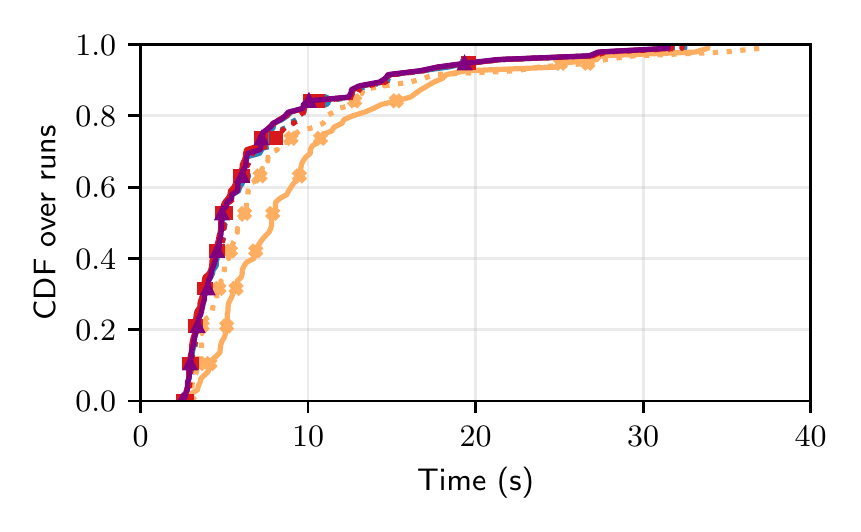}
		\caption{Transfer time.}
		\label{fig:time_homo}
	\end{subfigure}
	\hfill
	\begin{subfigure}[b]{0.33\textwidth}
		\centering
		\includegraphics[width=\columnwidth]{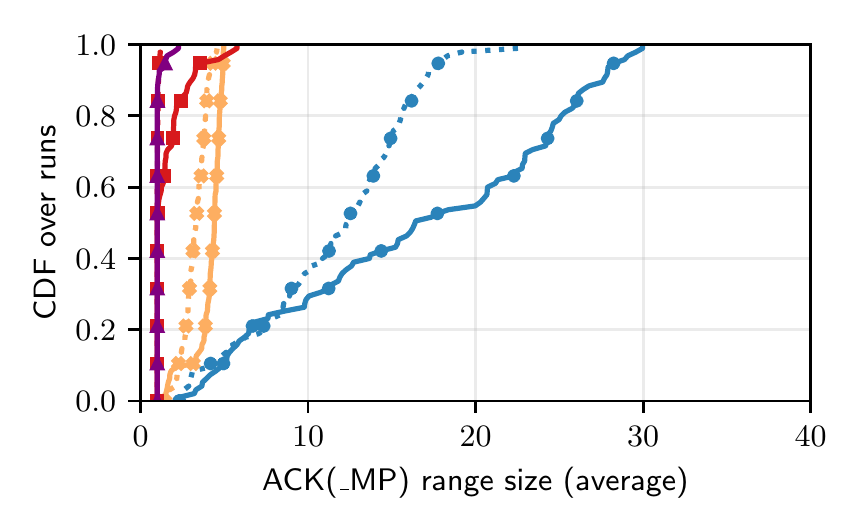}
		\caption{Range size advertised by the client.}
		\label{fig:ack_ranges_homo}
	\end{subfigure}
	\hfill
	\begin{subfigure}[b]{0.33\textwidth}
		\centering
		\includegraphics[width=\columnwidth]{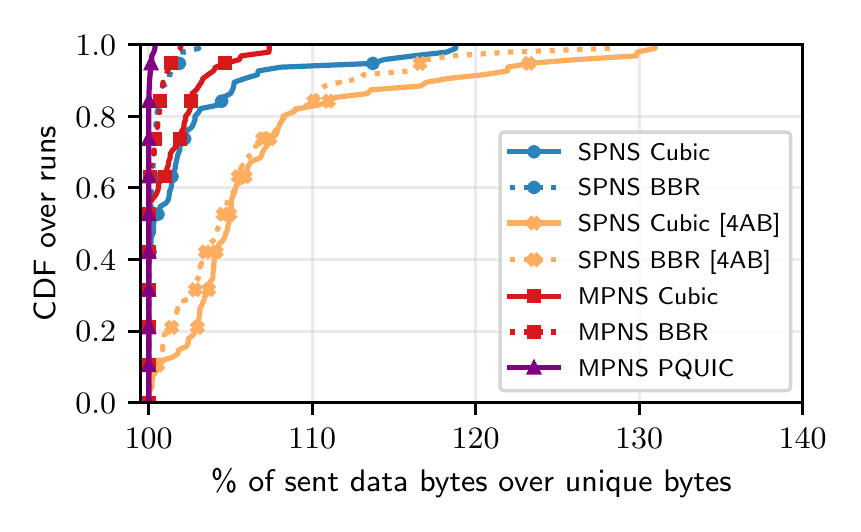}
		\caption{Data sent by the server.}
		\label{fig:stream_retrans_homo}
	\end{subfigure}
	\caption{Homogeneous experiments. The legend is common to all the figures.}
	\label{fig:homo}
\end{figure*}

Figure~\ref{fig:time_homo} shows the completion times for the 50 MB transfer.
As \picoquic supports both packet number space designs and two different congestion control schemes (Cubic and BBR), we evaluate each combination.
\pquic only supports \mpns with Cubic.
The transfer times are mostly dependent of the paths' bandwidth that differs between runs.
Still, we do not observe much performance difference between \spns and \mpns, which is expected.

However, having homogeneous paths does not prevent the receiver from observing reordering.
We extract the ranges advertised in the \texttt{ACK} (for \spns) and \texttt{ACK\_MP} frames (for \mpns) sent by the client.
Figure~\ref{fig:ack_ranges_homo} shows the average size of the \texttt{ACK(\_MP)} ranges over runs.
With \mpns, \texttt{ACK\_MP} frames often show a single range, as no reordering occurs within a path.
Still, a few packet losses due to buffer overflow might occur, leading to a few \mpns runs where the average range size is greater than 1.
With \spns, \texttt{ACK} frames can carry many ranges, even if there is no buffer loss.
Indeed, it can happen that the server pushes slightly more on a path than the other, hence adding queuing delay on that path, causing an initial reordering.
Afterwards, the reception of \texttt{ACK} frames with several ranges lets the server send a burst of packets on the less-loaded path.
\picoquic integrates pacing at the sender to limit this effect.


Yet, the high average range size of \spns raises concerns when the receiver wants to further limit the number of ACK Blocks (AB) it writes.
If there is reordering involving many packets, the client may never acknowledge a given packet, even if it was received.
To implement this situation, we lower the maximum number of ACK Blocks that an \texttt{ACK} frame can contain to 4, leading to at most 5 ranges within a given \texttt{ACK} frame.
While Figure~\ref{fig:ack_ranges_homo} confirms that the average range size is always lower than (but close to) 5, Figure~\ref{fig:time_homo} shows that such a strategy hinders the performance of the transfer, especially when using Cubic that tends to overflow buffers.
To explain this performance drop, we consider the \texttt{STREAM} frames the server sends.
Based on the data offset and length fields, we compute the number of retransmitted data bytes and make it relative to the transfer size (50 MB).
Figure~\ref{fig:stream_retrans_homo} indicates that there are indeed more data retransmission with \spns, and especially when the number of ACK Blocks that \texttt{ACK} frames can convey is limited.
Even in homogeneous networks, it is possible to be unable to acknowledge some packets due to the limited ACK blocks, leading to spurious retransmissions, decreased sending rate and lower overall performance.
Note that BBR is less affected by packet loss detection than Cubic, making the impact on its transfer performances lower.

\subsection{Heterogeneous 2-Path Experiments}\label{sec:hetero_2p}

\begin{table}
	\centering
	\caption{Parameter space for the 95 heterogeneous 2-path network scenarios.}
	\begin{tabular}{@{}c|c@{}}
		\toprule
		Factor & Value\\
		\midrule
		Total Bandwidth [Mbps] & 100 \\
		Total RTT [ms] & 200 \\
		Bandwidth Balance & [0.1; 0.9] \\
		RTT Balance & [0.1; 0.9] \\
		\bottomrule
	\end{tabular}
	\label{tab:hetero_param_space}
\end{table}

The previous experiments considered an "idealistic" case where all network paths share the same characteristics.
This situation rarely happens in practice with, e.g., Wi-Fi/LTE or terrestrial/satellite~\cite{zhang2006measurement}.
To investigate situations where paths have different properties, we consider the parameter space shown in Table~\ref{tab:hetero_param_space}.
We split the bandwidth and RTT between paths such as their sum is always the same.
For instance, if bandwidth balance is 0.9 and RTT balance 0.1, then the first path has 90 Mbps 20 ms RTT and the second path 10 Mbps 180 ms RTT.
As the theoretical bandwidth remains the same across all runs, this enables easier comparison between time performance.

\begin{figure*}
	\centering
	\begin{subfigure}[b]{0.33\textwidth}
		\centering
		\includegraphics[width=\columnwidth]{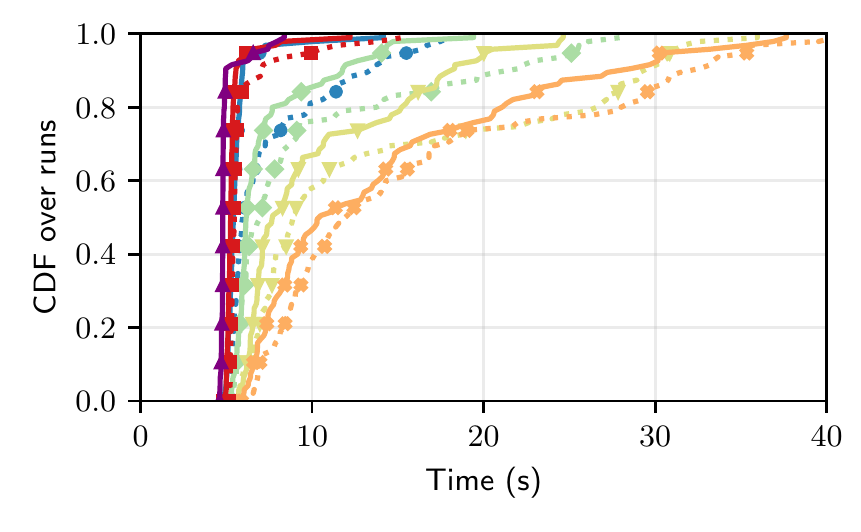}
		\caption{Transfer time.}
		\label{fig:time_hetero_before_fix}
	\end{subfigure}
	\hfill
	\begin{subfigure}[b]{0.33\textwidth}
		\centering
		\includegraphics[width=\columnwidth]{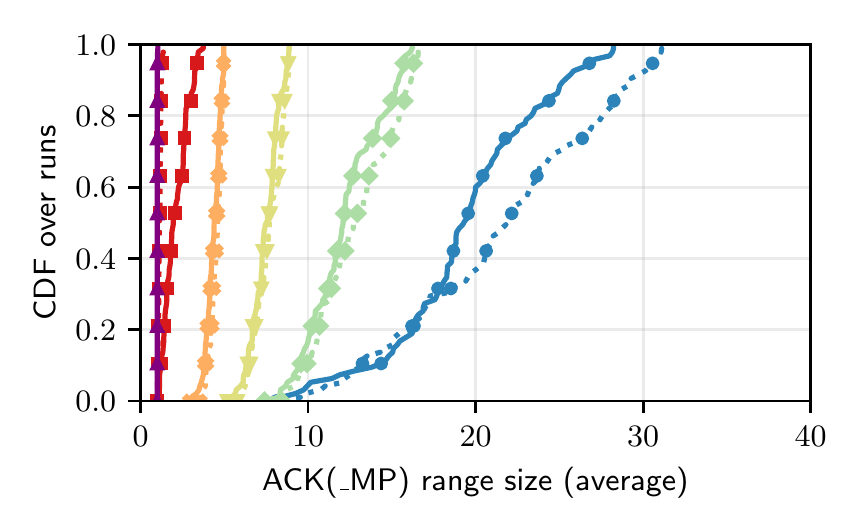}
		\caption{Range size advertised by the client.}
		\label{fig:ack_ranges_hetero_before_fix}
	\end{subfigure}
	\hfill
	\begin{subfigure}[b]{0.33\textwidth}
		\centering
		\includegraphics[width=\columnwidth]{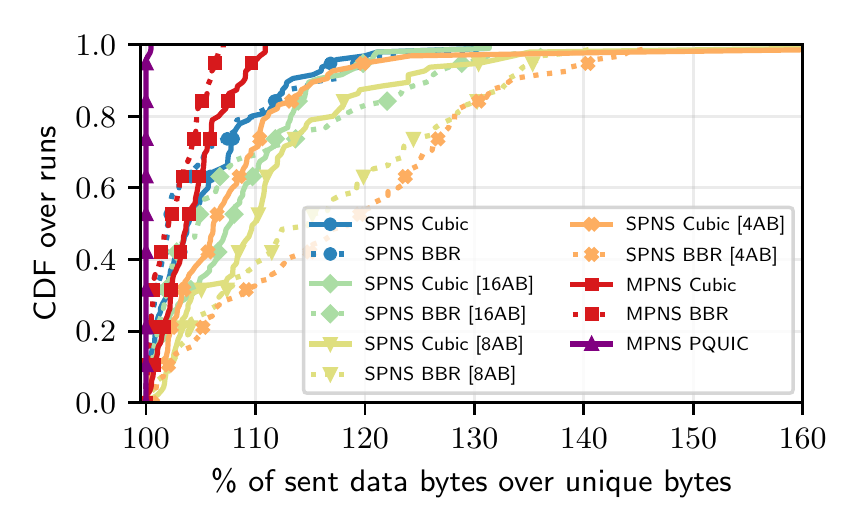}
		\caption{Data sent by the server.}
		\label{fig:stream_retrans_hetero_before_fix}
	\end{subfigure}
	\caption{Heterogeneous 2-path experiments with original \picoquic. The legend is common to all the figures.}
	\label{fig:hetero}
\end{figure*}

When using an upper limit of 32 ACK Blocks and considering Cubic, \picoquic with \spns keeps performance close to \picoquic with \mpns and \pquic, as depicted in Figure~\ref{fig:time_hetero_before_fix}.
However, like in homogeneous experiments, Figure~\ref{fig:ack_ranges_hetero_before_fix} shows that \picoquic with \spns triggers \texttt{ACK} frames advertising many ranges, sometimes hitting the receiver's ACK Blocks limit.
If we further limit the number of ACK Blocks being sent (to 16, 8 and 4) within an \texttt{ACK} frame, the transfer performance degrades because of spurious loss detection and increased data retransmission.
In particular, with \spns \picoquic using Cubic whose receiver does not advertise more than 4 ACK Blocks, there is a run where there are more than 70 \% of the data that get retransmitted, some piece of data being retransmitted up to 8 times.
Furthermore, the performance of \picoquic transfers using BBR are worse than the ones using Cubic.

\begin{figure*}
	\centering
	\begin{subfigure}[b]{0.33\textwidth}
		\centering
		\includegraphics[width=\columnwidth]{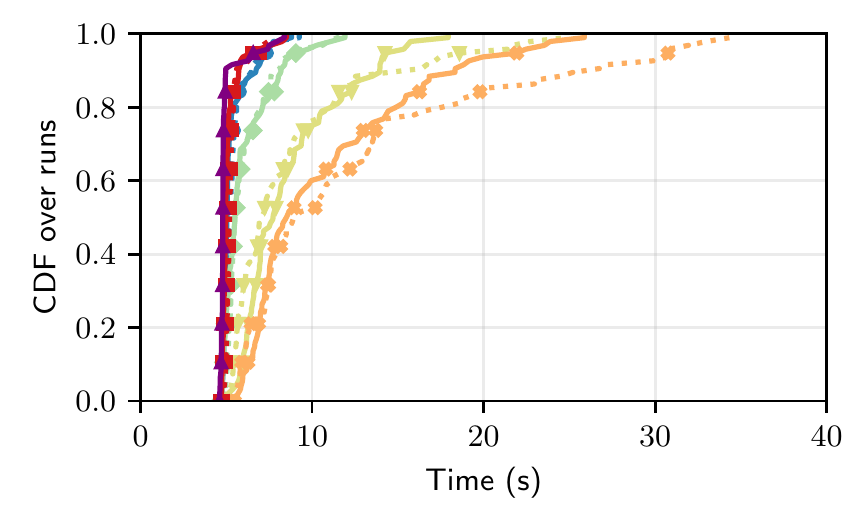}
		\caption{Transfer time.}
		\label{fig:time_hetero_after_fix}
	\end{subfigure}
	\hfill
	\begin{subfigure}[b]{0.33\textwidth}
		\centering
		\includegraphics[width=\columnwidth]{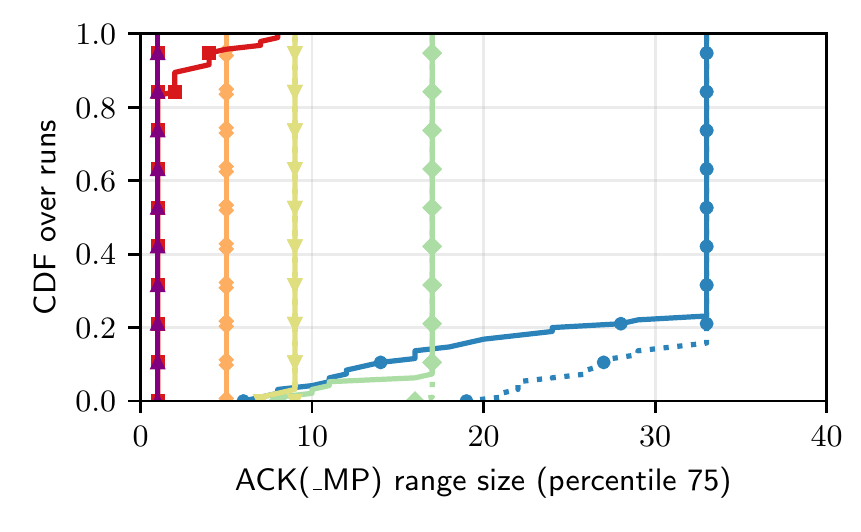}
		\caption{Range size advertised by the client.}
		\label{fig:ack_ranges_hetero_after_fix}
	\end{subfigure}
	\hfill
	\begin{subfigure}[b]{0.33\textwidth}
		\centering
		\includegraphics[width=\columnwidth]{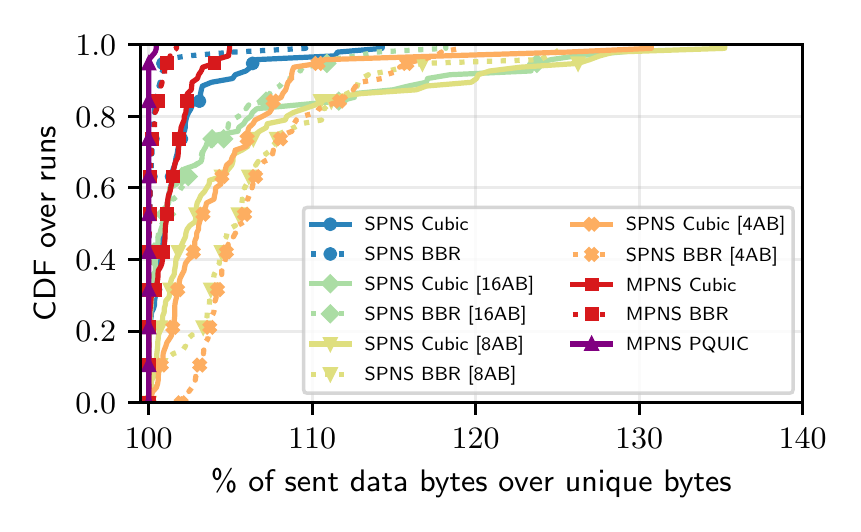}
		\caption{Data sent by the server.}
		\label{fig:stream_retrans_hetero_after_fix}
	\end{subfigure}
	\caption{Heterogeneous 2-path experiments with fixed \picoquic. The legend is common to all the figures.}
	\label{fig:hetero_fixed}
\end{figure*}

\begin{figure}
	\centering
	\includegraphics[width=.66\columnwidth]{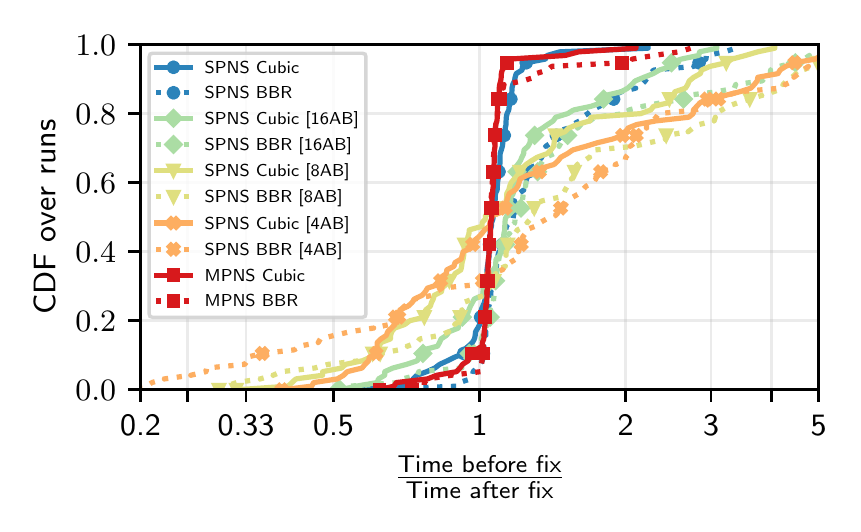}
	\caption{A ratio greater than 1 means that fixed \picoquic is faster than original \picoquic. Note the logarithmic x scale.}
	\label{fig:ratio_time}
\end{figure}

We contacted the \picoquic's implementer to report these results.
He was aware of the two aforementioned issues.
First, for the BBR performance, it was related to the \texttt{ACK} frame scheduling by the data receiver.
Under some circumstances, a specific range was always sent on the slow path.
Then, a later range arrived first on the fast path, affecting the path's RTT estimate and loss detection algorithm based on RACK~\cite{rfc8985}.
The implemented fix consists in duplicating \texttt{ACK(\_MP)} frames on both paths.
Second, regarding limited ACK Blocks, the implementation's behavior was to acknowledge the largest number ranges first.
In case of large packet reordering, it may happen that packets got either lately or never acknowledged, leading to spurious data retransmissions.
To address that, \picoquic now includes an heuristic placing the lowest ranges first, knowing the maximum number of ranges it wants to advertise.

\begin{figure}
	\centering
	\includegraphics[width=.66\columnwidth]{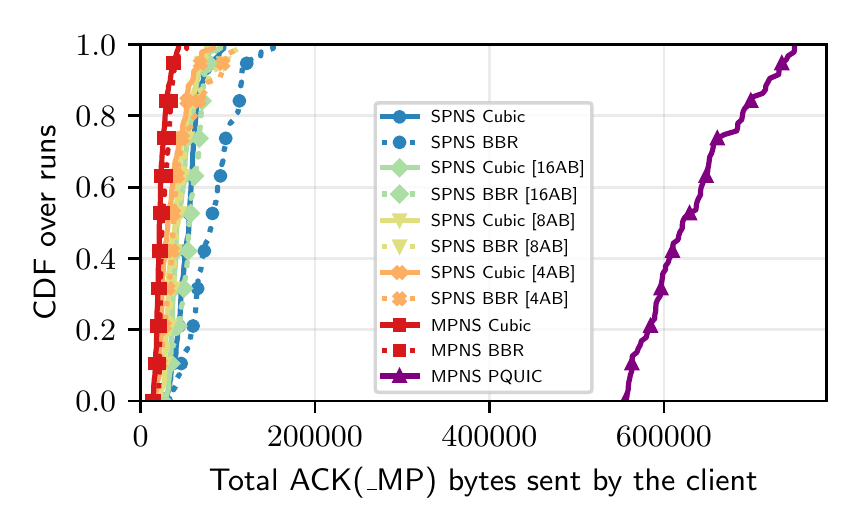}
	\caption{Total number of bytes of acknowledgment frames.}
	\label{fig:ack_length}
\end{figure}

\begin{figure*}
	\centering
	\begin{subfigure}[b]{0.33\textwidth}
		\centering
		\includegraphics[width=\columnwidth]{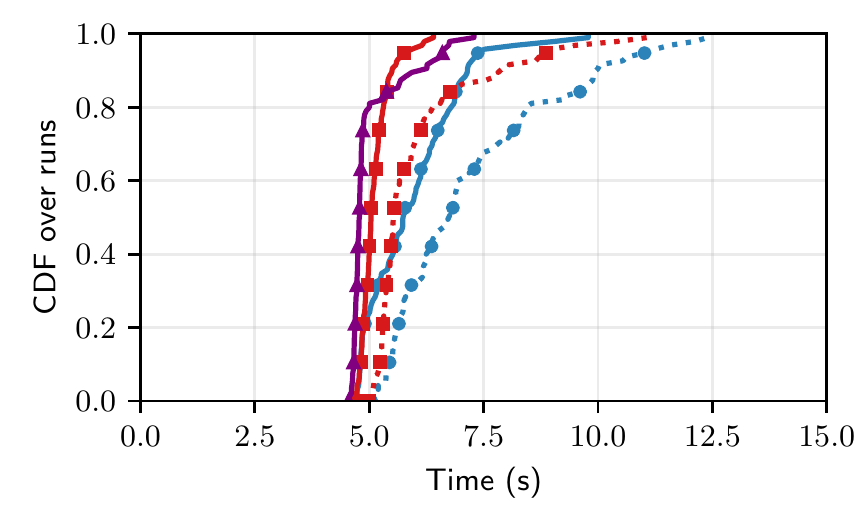}
		\caption{Transfer time.}
		\label{fig:time_hetero_3p}
	\end{subfigure}
	\hfill
	\begin{subfigure}[b]{0.33\textwidth}
		\centering
		\includegraphics[width=\columnwidth]{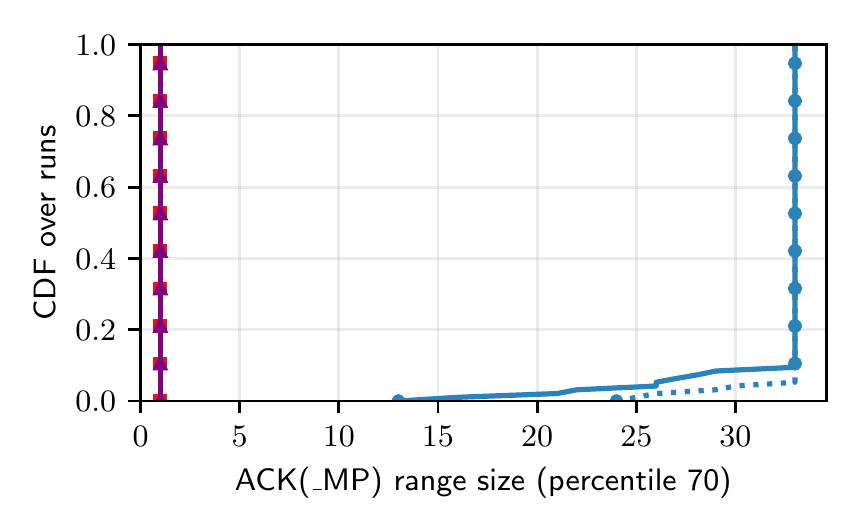}
		\caption{Range size advertised by the client.}
		\label{fig:ack_ranges_hetero_3p}
	\end{subfigure}
	\hfill
	\begin{subfigure}[b]{0.33\textwidth}
		\centering
		\includegraphics[width=\columnwidth]{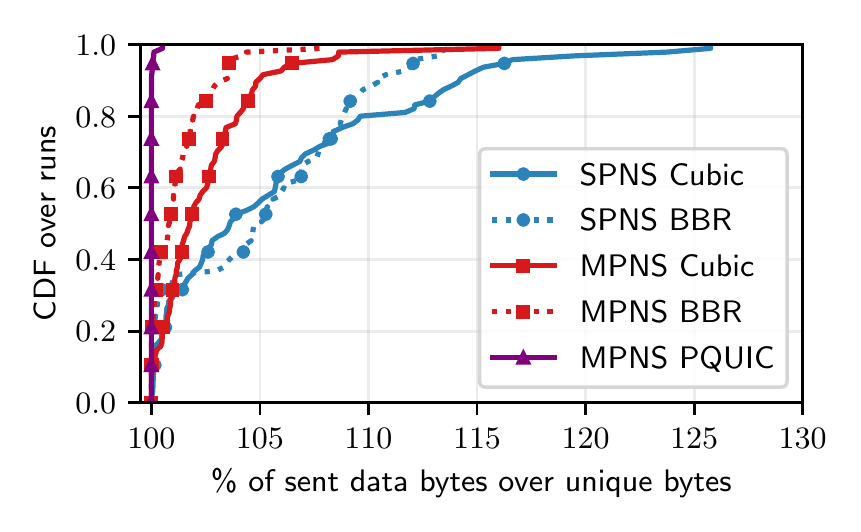}
		\caption{Data sent by the server.}
		\label{fig:stream_retrans_hetero_3p}
	\end{subfigure}
	\caption{Heterogeneous 3-path experiments with fixed \picoquic. The legend is common to all figures.}
	\label{fig:hetero_3p}
\end{figure*}

In the remaining experiments, we consider a version of \picoquic integrating these fixes\footnote{Commit \texttt{9eacfff}.}.
We rerun the previous 2-path heterogeneous experiments and present the results in Figure~\ref{fig:hetero_fixed}.
To better highlight the impacts of those fixes, Figure~\ref{fig:ratio_time} shows the ratio of the transfer time between the original \picoquic and the fixed one on the same run.
On the one hand, the performance of \picoquic using BBR considerably improved.
Indeed, BBR is very sensitive to the path's latency, and duplicating \texttt{ACK} frames on both paths makes these estimates more stable.
This ACK duplication also benefits to Cubic runs and the \mpns variant.
On the other hand, changing the range selection strategy provided mixed results.
As illustrated in Figure~\ref{fig:ack_ranges_hetero_after_fix}, in nearly all \spns runs, at least 25\% of the sent \texttt{ACK} frames hit the implementation limit of ACK Blocks.
While acknowledging lowest ranges first brings benefits in some network scenarios, it provides worse results in others.
If the receiver does not provide all information it has, it can trigger sub-optimal decisions at sender's side and increases the amount of retransmitted data.

Interestingly, some runs of \picoquic with \mpns using Cubic show data retransmissions and \texttt{ACK\_MP} frames with several ACK ranges while \pquic does not (like \picoquic BBR).
Two elements might explain this result.
First, the implementations of Cubic are slightly different, and \picoquic's one uses the estimated path's RTT to determine whether it should exit the slow-start phase or not.
This might make the \picoquic's Cubic slightly more aggressive than \pquic's one.
Second, as depicted in Figure~\ref{fig:ack_length}, the \pquic receiver acknowledges data more aggressively than the \picoquic one.
Indeed, the \pquic receiver sends an \texttt{ACK\_MP} frame for both paths as soon as two new packets arrive for a given path.
In the case of \picoquic, while its initial limit is also set to 2, it supports the \texttt{ACK\_FREQUENCY} frame~\cite{ietf-quic-ack-frequency-01} and requests its peer to not trigger immediate acknowledgments if it detects packet number reordering.
It can then wait a few milliseconds between subsequent packets containing \texttt{ACK(\_MP)} frames.
However, during that time lap, several tens of packets from the server can have been received.
Also note that while the minimum size of an \texttt{ACK\_MP} frame is larger than a regular \texttt{ACK} one, \picoquic with \mpns manages to send less acknowledgment-related bytes than \picoquic with \spns.
Smaller ranges and less out-of-order triggered acknowledgments explain this result.

\subsection{Heterogeneous 3-Path Experiments}

\begin{table}
	\centering
	\caption{Parameter space for the 95 heterogeneous 3-path network scenarios.}
	\begin{tabular}{@{}c|c@{}}
		\toprule
		Factor & Value\\
		\midrule
		Total Bandwidth [Mbps] & 100 \\
		Total RTT [ms] & 300 \\
		Bandwidth Path Weight & [0.1; 0.9] \\
		RTT Path Weight & [0.1; 0.9] \\
		\bottomrule
	\end{tabular}
	\label{tab:hetero_3p_param_space}
\end{table}

When considering multipath, one usually starts with two network paths.
Still, there are cases where more than two network paths are simultaneously available, such as dual-stacked Wi-Fi and LTE.
To cover them, we consider 3-path scenarios covering the parameter space depicted in Table~\ref{tab:hetero_3p_param_space}.
Unlike in previous subsection that explored a 2-dimensional space, each path has a weight for the budget bandwidth and RTT, leading to 6 varying parameters.
As an example with the RTT, if the first path has a weight of 0.5, the second path 0.25 and the third path 0.75, the RTT of each path will respectively be 100 ms, 50 ms and 150 ms.

In the remaining experiments, we keep \pquic and the fixed version of \picoquic.
Figure~\ref{fig:hetero_3p} shows that when three heterogeneous paths are simultaneously used, 33 ranges inside \texttt{ACK} frames are often not sufficient with \spns to provide a full view of the received packets to the sender.
Indeed, as emphasized by Figure~\ref{fig:ack_ranges_hetero_3p}, in nearly all our runs, at least 30\% of all the \texttt{ACK} frames sent by the client hit the implementation limit.
This lack of information triggers more spurious loss detection events and retransmissions (Figure~\ref{fig:stream_retrans_hetero_3p}), leading to lower performance compared to \mpns (Figure~\ref{fig:time_hetero_3p}).

In such networks, it is still possible to make the \spns design work as well as the \mpns one by increasing or removing the maximum number of ACK Blocks that an implementation is willing to write in a single \texttt{ACK} frame.
However, when using \spns, this makes multipath more dependent on the receiver's implementation than with \mpns.
Note that increasing the acknowledgment rate of \picoquic enables it to get results closer to the \pquic's one while keeping the same trend between the different designs (figure not shown due to space constraints).

\section{Discussion and Next Steps}\label{sec:discussion}

In this paper, we evaluated the impact of the amount of packet number spaces on a bulk transfer.
While this scenario does not cover all the possible usages of multipath, it assesses its bandwidth aggregation feature.
To this end, we showed that both designs can address this need.
However, our experiments pointed out that when using the \spns design, the performance of Multipath QUIC is more dependent on the receiver. 

As already discussed by Huitema~\cite{huitema-pns}, the final decision of using one or several packet number spaces will be a trade-off.
On the one hand, a single packet number space provides support for zero-length Connection IDs, fewer protocol additions and no change to the nonce computation.
On the other hand, multiple packet number spaces allow implementations to keep the current loss detection algorithms per-path, do not add complexity when generating \texttt{ACK}s (including their frequency) and make the performance more resilient to the receiver's choices.
Knowing that the inclusion of multiple paths adds new dimensions to cope with (packet scheduling, path management), it might be wise to avoid adding complexity to these algorithms when using multipath transport protocols.


\paragraph{Artifacts.}
The evaluation/measurement/analysis scripts will be released in the future.
If you are already interested by them now, please contact the author.



{ \balance
{
	
	%
	\bibliographystyle{ACM-Reference-Format}
	\bibliography{biblio}
}
}

\end{document}